\documentclass[10pt]{article}
\renewcommand{\Re}{\mathop{\rm Re }\nolimits}

\def\refname{References}

\def\thebibliography#1{{\section*{\refname}}\list
 {\arabic{enumi}.}{\settowidth\labelwidth{[#1]}\leftmargin\labelwidth
 \advance\leftmargin\labelsep
 \usecounter{enumi}}
 \def\newblock{\hskip .11em plus .33em minus .07em}
 \sloppy\clubpenalty4000\widowpenalty4000
 \sfcode`\.=1000\relax}

\begin{document}

\begin{center}
\large \bf  SUPERHEAVY PARTICLES IN FRIEDMANN \\
COSMOLOGY AND THE DARK MATTER PROBLEM
\end{center}

\begin{center}
{\large A. A. GRIB}\footnote{ E-mail: grib@friedman.usr.lgu.spb.su }  \\
{\it \small   A.Friedmann Laboratory for Theoretical Physics,  \\
     30/32 Griboedov can, St.Petersburg, 191023, Russia    }
\end{center}

\begin{center}
{\large YU. V. PAVLOV}\footnote{ E-mail: pavlov@ipme.ru}    \\
{\it \small  Institute of Mechanical Engineering,
         Russian Academy of Sciences, \\
   61 Bolshoy, V.O., St.Petersburg, 199178, Russia}
\end{center}

\begin{abstract}
\noindent
   The model of creation of observable particles and particles of the dark
matter, considered to be superheavy particles, due to particle creation by
the gravitational field of the Friedmann model of the early Universe is
given. Estimates on the parameters of the model leading to observable
values of the baryon number of the Universe and the dark matter density
are made.
                   \end{abstract}
\vspace{5mm}
    It is well known${}^{1,2}$ that creation of superheavy particles
with the mass of the order of Grand Unification
$M_X \approx 10^{14} - 10^{15} $\, Gev with consequent decay on quarks and
leptons with baryon charge and CP -- nonconservation is sufficient for
explanation of the observable baryonic charge of the Universe.
   Recently in papers${}^{3,4}$ the possibility of explanation of
experimental facts on observation of cosmic ray particles with the energy
higher than the Greizen-Zatsepin-Kuzmin limit was discussed.
    The proposal is to consider the decay of superheavy particles
with the mass of the order $M_X$\,.
   One can even consider the hypothesis that all dark matter consists
of neutral $X $~-- particles with very low density.
   So it seems interesting to construct a model in which all matter as
the observable one as the dark matter arise due to creation of
$X $ -- particles by the gravitational field of the early Friedmann
Universe.    Gravitational field of the expanding Universe creates from
vacuum particle and antiparticle pairs of $X $~-- bosons.
    In papers,${}^{3,4}$ however, it was shown that if superheavy
particles were stable, then the Universe will collapse very quickly.
    That is why as it was claimed in Refs. 1,2
these particles must decay with baryon charge and CP nonconservation.
    Then one has the problem how these particles can survive  up to modern
time and lead to observable cosmic rays effects?
   However, considering their decay in full analogy with theory of
$K^0 $~-- mesons one must speak about decaying  $X_1^0 $ ,
$X_2^0 $~-- measons, which for simplicity are considered by us as conformal
scalar particles, with different life times and per cent content in the
initial mixture of $X $ and $\bar{X} $ bosons.

    So the problem is in the numerical estimate of the parameters of the
effective Hamiltonian leading to the observable data.
    Short living $ X^0 $~-- bosons decay on quarks and leptons in time
close to singularity, long living $ X^0 $ -- bosons exist today as the
dark matter.     Here we shall give this estimate. We shall not suppose
any explicit realization (for example $ SU(5) $ etc) of the Grand
Unification, because any such realization with baryon charge and
CP nonconservation must lead to our effective Hamiltonian.
   We also don't discuss the important problem of spontaneous breaking of
the Grand Unification symmetry, surely influencing the difference of life
times of particles of our model.
  We also neglect the effect of CPT -- breaking in the expanding Universe,
which can be important in the era of particle creation.${}^{5}$

    At first let us remind some important results on creation of
$ X $ -- particle pairs in the early Friedmann Universe. Pairs of conformal
scalar massive $X $ -- particles are created by the external gravitational
field, the source of which is radiation with nonzero entropy and
$ p=\varepsilon /3 $ equation of state.
   We don't discuss here the origin of this entropy (inflation or some
other model) as well as creation of pairs from radiation.
   Particles are created by gravitation at the Compton time
$ t \sim M^{-1} $ and for $ t \gg M^{-1} $ one has nonrelativistic
gas of created particles with the energy density calculated for the
radiation dominated Friedmann model${}^{1}$
$a(t)=a_0 \, t^{1/2} $:
\begin{equation}
\varepsilon^{(0)}=2b^{(0)}\,M(M/t)^{3/2} \,,
\end{equation}
    where $b^{(0)}=5.3\cdot 10^{-4} $.
Total number of created particles in the Lagrange volume is
\begin{equation}
N=n^{(0)}(t)\,a^3(t)=b^{(0)}\,M^{3/2}\,a_0^3 \ .
\end{equation}
    In spite of the cosmological order of the number of created
$X $ -- particles
($ N \sim 10^{80} $ for $ M_X \sim 10^{15} $\,Gev, see Ref. 1.)
their back reaction on the background metric is small in the sense that
the background metric cannot arise due to these particles:${}^{1}$
\begin{equation}
\frac{\varepsilon_{part}}{\varepsilon_{bground}} \sim
\frac{G}{t^2}=\left[\frac{1/t}{M_{Pl}}\right]^2 \ll 1  \,,
\ \ \ \mbox{for} \ \ \ t_{Pl} \ll t \ll \frac{1}{M} \,.
\end{equation}
    However for ${t \gg M^{-1}} $ there is an era of going from the
radiation dominated model to the dust model of superheavy particles for
$ \varepsilon_{bground}~\approx~\varepsilon^{(0)} $\,,
\begin{equation}
t_X\approx \left(\frac{3}{64 \pi \, b^{(0)}}\right)^2\,
\left(\frac{M_{Pl}}{M_X}\right)^4\,\frac{1}{M_X}  \,.
\end{equation}
    If $M_X \sim 10^{14} $\,Gev,  $\ t_X \sim 10^{-15} $\,sec, if
$M_X \sim 10^{13} $\,Gev -- $t_X \sim 10^{-10} $\,sec.
    So the life time of short living $X $ -- mesons must be smaller then
$t_X $\,. It is evident that if all created  $X $ -- particles were stable,
than the closed Friedmann model could quickly collapse, while all other
models are strongly different from the observable Universe.
   Let us define $d $ -- the permitted part of long living
$X $ -- mesons --- from the condition: on the moment of
recombination $t_{rec} $ in the observable Universe one has
$$
d\,\varepsilon_X(t_{rec}) =\varepsilon_{crit}(t_{rec})  \,,
$$
\begin{equation}
d=\frac{3}{64 \pi \, b^{(0)}}\left(\frac{M_{Pl}}{M_X}\right)^2\,
\frac{1}{\sqrt{M_X\,t_{rec}}}\, .
\label{d}
\end{equation}
For $M_X=10^{13} - 10^{14} $\,Gev one has
$d \approx 10^{-12} - 10^{-14} $\,.
    Using the estimate for the velocity of change of the concentration of
long living superheavy particles${}^{6}$
$|\dot{n}_x| \sim 10^{-42}\, \mbox{cm}^{-3}\,\mbox{sec}^{-1} $,
and taking the life time $\tau_X $ of long living particles as
$2\cdot 10^{22} $\,sec we obtain concentration
$n_X \approx 2\cdot 10^{-20} \,\mbox{cm}^{-3} $ at the modern epoch,
corresponding to the critical density for $M_X=10^{14} $\,Gev\,.

    Now let us construct the toy model which can give: \ \
a) short living $X $ -- mesons decay in time
   $\tau_q < 10^{-15} $\,sec, (more wishful is
   $\tau_q \sim 10^{-38} - 10^{-35} $\,sec),
   long living mesons decay with $\tau_l > t_U \approx 10^{18}$\,sec
   \  ($t_U $~--~age of the Universe).        \ \ \
b) one has small $ d \sim 10^{-14} - 10^{-12} $ part of long living
   $X $ -- mesons, forming the dark matter.

   Baryon charge nonconservation with CP -- nonconservation in full analogy
with the $K^0 $ -- meson theory with nonconserved hypercharge and
CP -- nonconservation leads to the effective Hamiltonian of the decaying
$X, \bar{X} $ -- mesons with nonhermitean matrix.

   The matrix of the effective Hamiltonian is in standard
notations${}^{7}$
\begin{equation}
H=     \left(
\begin{array}{cc}
H_{11}  & H_{12}   \\
H_{21}  & H_{22} \\
\end{array}        \right) .
\label{H}
\end{equation}
    Let $H_{11}\! =\! H_{22}$
(due to $CPT$-invariance).
Denote
$\ \varepsilon=(\sqrt{\vphantom{ }H_{12}} - \sqrt{H_{21}}\,)\, / \,
(\sqrt{H_{12}} + \sqrt{H_{21}} \, )$.
    The eigenvalues $\lambda_{1,2} $ and eigenvectors
$|\,\Psi_{1,2}\!> $  of matrix $H$ are
    \begin{equation}
\lambda_{1,2} = H_{11} \pm \frac{H_{12}+H_{21}}{2} \,
\frac{1-\varepsilon^2}{1+\varepsilon^2} \,,
\end{equation}
    \begin{equation}
|\,\Psi_{1,2}\!>=\frac{1}{\sqrt{2\,(1+|\varepsilon |^2)}}\,
\biggl[(1+\varepsilon) \,|\,1\!> \pm \,(1- \varepsilon) \, |\,2\!>\biggr].
\end{equation}

         In particular
\begin{equation}
H=     \left(
\begin{array}{cc}
E-\frac{i}{4}\left(\tau_q^{-1} +\tau_l^{-1}\right)
  &
\frac{1+\varepsilon}{1-\varepsilon}
\left[A-\frac{i}{4}\left(\tau_q^{-1} -\tau_l^{-1}\right)\right]
 \\  & \\
\frac{1-\varepsilon}{1+\varepsilon}
\left[A-\frac{i}{4}\left(\tau_q^{-1} -\tau_l^{-1}\right)\right]
 &
E-\frac{i}{4}\left(\tau_q^{-1} +\tau_l^{-1}\right) \\
\end{array}        \right) .
\label{HM}
\end{equation}

Then the state $|\,\Psi_1 \!> $ describes
short living particles with the life time
$ \ \tau_q \ $ and mass $E+A$.
The state $\ |\,\Psi_2 \!> $ is the state of long living particles
with life time $ \tau_l \ $  and mass $E-A$.
    Here $A$ is the arbitrary parameter $-E<A<E$  and it can be zero,
$E=M_X$.

If $ d = 1 - |\!<\!\Psi_1\,|\,\Psi_2\!>\!|^2 = 1 - |2 \Re
\varepsilon /(1+|\varepsilon |^2)|^2 $ is the relative part of
long living particles and $\varepsilon $ is real, then $
\varepsilon=(1-\sqrt{d}\,)/\sqrt{1-d}$ \ ($\varepsilon \sim
1-10^{-7} $ for $M_X \sim 10^{14}$\,Gev). So one has a typical
example of "fine tuning" in order to obtain the desired result.

Taking $\tau_q=10^{-35}$ sec and $\tau_l=2\cdot 10^{22}$ sec, one obtains
that for the Hermitean part $(H+H^{+})/2$ of $H$ nondiagonal
$CP$-noninvariant term equal to
$ i\, (\tau_q^{-1} - \tau_l^{-1})\, \varepsilon / (1-\varepsilon^2)/2 $
is of the order of Planckean mass $10^{19}$ Gev.
\vspace{3mm}     \\
{\bf Acknowledgments}                     \\
This work was supported by Min. of Education of Russia, grant E00-3-163.

              { \fontsize{9pt}{11pt} \selectfont
    }
\end{document}